\documentclass[a4paper,aps,twocolumn,showpacs,showkeys,nofootinbib,preprintnumbers,superscriptaddress,amsmath,amssymb,amsfonts]{revtex4-1}
\usepackage{graphicx}
\usepackage{mathrsfs} 
\usepackage{amsmath}
\usepackage{bm,natbib}
\usepackage[utf8]{inputenc}
\usepackage[T1]{fontenc}
\usepackage[english]{babel}
\usepackage{amssymb}
\usepackage{amsfonts}
\usepackage{epsfig}
\usepackage{colordvi}
\newcommand{\Lagr}{\mathcal{L}}
\newcommand{\G}{\mathcal{G}}
\usepackage{psfrag}
\usepackage{color}
\usepackage{dcolumn}
\usepackage{multirow}
\usepackage{hyperref}
\usepackage{epstopdf}
\usepackage{bm}
\usepackage{times}
\hypersetup{
  colorlinks=true,        
  linkcolor=blue,         
  citecolor=magenta,      
}

\begin{document}

\title{Noether Symmetries and Quantum Cosmology in Extended Teleparallel Gravity}

\author{Francesco Bajardi}
\email{francesco.bajardi@unina.it}
\affiliation{Department of Physics ``E. Pancini'', University of Naples ``Federico II'', Naples, Italy.}
\affiliation{INFN Sez. di Napoli, Compl. Univ. di Monte S. Angelo, Edificio G, Via Cinthia, I-80126, Naples, Italy.}

\author{Salvatore Capozziello}
\email{capozziello@na.infn.it}
\affiliation{Department of Physics ``E. Pancini'', University of Naples ``Federico II'', Naples, Italy.}
\affiliation{INFN Sez. di Napoli, Compl. Univ. di Monte S. Angelo, Edificio G, Via	Cinthia, I-80126, Naples, Italy.}
\affiliation{Scuola Superiore Meridionale, Largo San Marcellino 10, I-80138, Naples, Italy.}
\affiliation{Tomsk State Pedagogical University, ul. Kievskaya, 60, 634061 Tomsk, Russia.}

\date{\today}
\begin{abstract}
We apply the Noether Symmetry Approach to point-like teleparallel Lagrangians in view to derive minisuperspaces suitable for Quantum Cosmology. Adopting the Arnowitt--Deser--Misner formalism, we find out related  Wave Functions of the Universe. Specifically, by means of  appropriate changes of variables suggested by the existence of Noether symmetries, it is possible to obtain the cosmological Hamiltonians whose solutions are classical trajectories interpretable as observable universes.

\end{abstract}

\maketitle

\section{Introduction}
\label{sec:uno}
General Relativity (GR) is considered the best accepted theory describing gravity. Despite its successes, related to the recent observational discovery of gravitational waves and  black holes, after more than one hundred years from its formulation,  the theory presents many theoretical and experimental issues, both at low and high energy scales, which need to be addressed. At infrared scales, there are problems related to  the standard cosmological model and the lack of explanation, at fundamental scales,  of dark energy and dark matter.  At ultraviolet scales,  GR cannot be renormalized as the other field theories, and in general its formalism cannot be adapted to Quantum Field Theory. Quantum theories of fields, in fact, deal with a standard Minkowskian background and fields can be treated separately from spacetime. In GR, the field turns out to be the background itself, and this represents a severe obstacle towards the construction of a self-consistent Quantum Gravity. The impossibility of fixing the ultraviolet divergences is also due to the fact that GR is a diffeomorphism-invariant, covariant theory which cannot be treated under the standard Yang-Mills formalism on a fixed background. 
 
Considering this state of art, effective theories have been proposed to cure GR shortcomings at large and microscopic scales. For example, at astrophysical and cosmological scales, extending the Einstein-Hilbert action through functions of the Ricci scalar - $f(R)$ theory - or through other curvature invariants might be a good starting point in order to address dark side problems, as shown \emph{e.g.} in \cite{Capozziello:2011et, Capozziello:2007ec, Capozziello:2012ie, Nojiri:2006ri, Basilakos:2011rx, Capozziello:2013wha, Capozziello:2019cav, Novikov:2016hrc}.

On the other hand, in order to consider gravity under the same standard as other interactions, it is necessary to propose a quantization approach and describe gravitation by a gauge formalism. One of the first quantization schemes was introduced by Arnowitt, Deser and Misner (ADM) which constructed a formalism leading to a canonical quantization. Further details and applications of the ADM formalism can be found in \cite{Bajardi:2020fxh, Hawking:1995fd, Kuchar:1976yw, Ashtekar:1987gu}. This procedure  is based on the concept of an infinite-dimensional superspace constructed on the  3D- spatial metrics. Once  a Super-Hamiltonian is defined on it, it is possible to define a {\it geometrodynamics} to fix the evolution of these 3-metrics. 
 
The problem is extremely difficult to be handled from a mathematical point of view, nevertheless, in 1983 J.B. Hartle \cite{Hartle:1983ai}, using the Wentzel-Kramers-Brillouin approximation, showed that it is possible to restrict the superspace to finite-dimensional minisuperspaces where the SuperHamiltonian can be quantized and a Wave Function of the Universe can be  analytically found as the solution of the Wheeler-De Witt (WdW) equation. In particular,  Hartle formulated a  criterion by which the Wave Function of the Universe, showing correlations and  oscillating behaviors in the minisuperspace, gives rise to classical trajectories representing observable universes.

The Wave Function of the Universe  is only related to the probability amplitude to obtain given configurations (see, e.g.  \cite{Vilenkin:1988yd}) but it does not give  the full information on  probability. The interpretation of the Wave Function has been discussed for many years and it is still not completely clear. For instance, in its pioneering work, Everett proposed the so called \emph{Many Worlds Interpretation of Quantum Mechanics} \cite{Everett}. In this framework,  the Wave Function of the Universe  acquires a probabilistic meaning. According to such an interpretation, all possible results of quantum measurements are simultaneously realized in different universes without, therefore, a Wave Function collapse. In other words, all the information is brought by the Schr\"odinger equation (in this case by the WdW equation) and the wave function collapse giving real values of the observables (as in the Copenhagen picture) is not required. The weak point of the approach is understanding the role of experiments which would be able to verify the consistency of measurements.
 
 Another interpretation was provided by Hawking, who stated  that the Wave Function is related to the probability amplitude for the early universe to develop towards our classical universe. 
 
Despite these difficulties on the interpretation, the canonical quantization scheme is useful in Quantum Cosmology since, thanks to the Wave Function, it is possible to recover  classical trajectories representing  observable universes.  In other words, the Quantum Cosmology application revealed more useful than the full theory. 
 
However, the canonical quantization procedure does not allow  to obtain a full renormalizable theory of gravity  to be dealt under the standard of Quantum Field Theory. A first step aimed at solving such a problem, is to treat gravity as a gauge theory. As pointed out in Sec. \eqref{TG}, it turns out that the gravitational interaction can be seen as a gauge theory of the local translation group and the corresponding Lagrangian turns out to be equivalent to that of GR up to a boundary term \cite{Gonzalez:2011dr, Jamil:2012ck, Cai:2015emx, Maluf:1994ji, Cai:2015emx}. This theory is the so called \emph{Teleparallel Equivalent of General Relativity} (TEGR) proposed by Einstein himself some years later the publication of GR \cite{Pereira}. According to some authors, quantizing TEGR \cite{Ming} should be a realistic approach to realize a full Quantum Gravity theory.  

In this perspective, Quantum Cosmology related to TEGR and its extensions could be a useful exercise towards a comprehensive quantum theory of gravity. This paper is devoted to this aim. In particular we want to show that quantizing minisuperspaces derived from TEGR and its extensions gives realistic cosmological models. The main role in this study is played by the symmetries that, if exist for these  minisuperspaces, give a selection rule related to the Hartle criterion for classical  trajectories, as formerly demonstrated in \cite{Capozziello:1999xr, Capozziello:2012hm}. 

In the next sections we briefly overview some basic aspects of TEGR and its applications. Then we apply the Noether symmetry approach (see  \emph{e.g.} \cite{Capozziello:1998nd, Paliathanasis:2011jq, Capozziello:2008ch, Capozziello:2012hm, Capozziello:1999xs, Bajardi:2020xfj, Urban:2020lfk, Bajardi:2020mdp, Kucukakca:2013mya, Gecim:2017hmn, Kucukakca:2014vja} ) to some extended teleparallel Lagrangians containing  functions of  torsion scalar,  teleparallel equivalent Gauss-Bonnet topological invariant, and  higher-order torsion terms. The purpose is to show that extended teleparallel models naturally exhibit Noether symmetries  allowing a straightforward quantization of the related minisuperspace models and then the possibility to recover observable universes.
 
The paper is organized as follows: in Sec. \ref{TG}, we discuss the main features and the basic foundations of Teleparallel Gravity (for reviews on the topic, see \cite{Cai:2015emx, Pereira, Arcos:2005ec}). In Secs. \ref{f(T)}, \ref{HO}, \ref{TGa} and \ref{Scalar}, we apply the Noether approach to different cosmological point-like Lagrangians. Then, thanks to an appropriate change of variables suggested by Noether's Theorem, we quantize the cosmological Hamiltonians and find the corresponding Wave Function of the Universe, solution of the WdW-Schr\"odinger-like equation. In Sec. \ref{concl},  we  discuss  results and  future perspectives. 

\section{Teleparallel Gravity\\}
\label{TG}

Canonical quantization of GR, reported in Appendix \ref{ADM},  does not fully address the problem of dealing with GR under the standard of Quantum Field Theory. Even assuming the ADM formalism, it is not possible to treat GR as a unitary gauge field theory like Electroweak interaction or Quantum Chromo Dynamics. Several shortcomings arise in the attempt to quantize a tensor field which is also the background of the theory. As said above, Quantum Field Theory instead, dealing with a fixed background, allows to  handle a quantum formalism by imposing  canonical commutation relations. In this picture,  many ultraviolet divergences cancel out by means of  Renormalization procedure. This latter, however, cannot be applied to GR, whose divergences cannot be straightforwardly regularized. 

Also for these reasons,  it is possible to consider  alternative theories  dealing with gravity as  a gauge theory of the translation group and describing the spacetime structure by torsion instead of curvature (see \cite{Cai:2015emx} for a conceptual discussion on this point). In what follows we recall some useful features related to  TEGR, considering results reported in \cite{Ferraro:2006jd, Ferraro:2008ey, Hammond:2002rm, Wu:2010mn, Maluf:2013gaa}.

Let us start by introducing the  \emph{tetrad fields} $h^a_\mu$,  used  to locally link the four-dimensional  spacetime manifold with its  tangent space, by means of the relation \footnote{Latin indexes label the tangent spacetime, while greek indexes are the coordinates labeling the standard spacetime.}
\begin{equation}
g_{\mu \nu} = h^a_\mu h^b_\nu \eta_{ab} \;.
\label{deftet}
\end{equation}
By means of the above relation, diffeomorphism transformations can be thought as translations in the locally flat tangent spacetime. In this way, the covariant derivative of a generic vector field $\phi^b$ can be written as:
\begin{equation}
D_a \phi^b = \partial_a \phi^b + \omega^b_{\,\,ac} \phi^c,
\end{equation}
where $\omega$ is the spin connection. For a tensor field with mixed indexes, both the spin connection and the Levi-Civita connection are involved, so that the covariant derivative of a rank-2 tensor $V^a_\nu$ turns out to be
\begin{equation}
\nabla_\mu V^a_\nu = \partial_\mu V^a_\nu + \omega^a_{b \mu} V^b_\nu + \Gamma^\alpha_{\mu \nu} V^a_\alpha \;.
\end{equation}
By requiring tetrads to satisfy the relation  $\nabla_\mu h^a_\nu=0$ and neglecting the spin connection, the so called Weitzenb\"ock connection arises:
\begin{equation}
\Gamma_{\mu \nu}^p = h^p_a \partial_\mu h^a_\nu \;.
\label{Weitz1}
\end{equation}
This assumption selects the class of frames with vanishing spin connection, though generally by means of "good tetrads" criteria it is possible to deal with both tetrads and spin connections \cite{Tamanini:2013xya}. The connection \eqref{Weitz1} is not symmetric with respect to the lowest indexes and the antisymmetric part, that is 
\begin{equation}
T_{\mu \nu}^p := 2 \Gamma_{[\mu \nu]}^p  \;,
\end{equation} 
defines the \emph{torsion tensor}. The contraction of the torsion tensor with a potential $S^{\mu \nu}_{\,\,\,\,\,p}$ defines the \emph{torsion scalar}:
\begin{eqnarray}
T = T_{p \mu \nu} S^{p \mu \nu} && \;\;\;\; S^{p \mu \nu} = K^{\mu \nu p}-g^{p \nu} T_{\,\,\,\,\,\,\, \sigma}^{\sigma \mu}+g^{p \mu} T_{\,\,\,\,\,\,\, \sigma}^{\sigma \nu}, \nonumber
\\
&& K^\nu_{p \mu} = \frac{1}{2} \left(T_{p \;\; \mu}^{\; \nu} + T_{\mu \;\; p}^{\; \nu} - T^\nu_{\; p \mu}\right) \;.
\end{eqnarray}
TEGR postulates the action
\begin{equation}
S = \int h \; T \; d^4x \;,
\label{teleparallel action0}
\end{equation}
to describe the gravitational interaction, where $h$ is the determinant of the tetrad fields. It is possible to show that the above action differs from the Einstein-Hilbert one only for a boundary term. Specifically,  the relation
\begin{equation}
R - \frac{2}{h} \partial_\mu (h T^{\nu \mu}_{\;\;\;\; \nu}) = -T,
\end{equation} 
holds. This means that the action \eqref{teleparallel action0} automatically yields the same field equations as GR. The field equations can be obtained by varying the action with respect to the tetrad fields, and read as
\begin{equation}
\partial_\sigma (h \; S_a^{\;\; p \sigma}) +h \; h^\lambda_a S^{\nu p}_{\;\;\;\; \mu} T^\mu_{\; \nu \lambda} = 0.
\label{teleparallel field equations}
\end{equation}
Similarly to GR extensions, TEGR action can be modified with the aim of solving the large-scale structure issues \cite{Cai:2015emx}. The simplest extension is given by the action containing a function of the torsion scalar, namely:
\begin{equation}
S = \int h f(T) d^4x \;.
\label{teleparallel action}
\end{equation}
As showed in \cite{Myrzakulov:2013hca, Fayaz:2014swa, Finch:2018gkh}, some particular functions are able to explain the cosmic accelerated  expansion and the structure formation without  introducing dark energy and dark matter. It is worth noticing that while teleparallel field equations are equivalent to those of GR, $f(R)$ and $f(T)$ theories are quite different. For instance, as the former leads to fourth-order equations of motion, the latter provides second-order equations. This difference, besides simplifying the dynamics of the $f(T)$ theory, has several implications in the cosmological framework. As an example, the further polarization modes carried by $f(R)$ gravitational waves, vanish in the $f(T)$ formalism, whose modes are the same as in standard TEGR \cite{Capozziello:2019klx, Bamba:2013ooa, Abedi:2017jqx, Capriolo1,Capriolo2}. Extended TEGR field equations read:
\begin{eqnarray}
&&\frac{1}{h} \partial_\mu(h \; h^p_a S_p^{\;\; \mu \nu}) f_T(T) - h^\lambda_a T^p_{\;\; \mu \lambda} S_p^{\;\; \nu \mu} f_T(T)+\nonumber 
\\
&&+ h^p_a S_p^{\;\; \mu \nu}(\partial_\mu T) f_{TT}(T) + \frac{1}{4} h^\nu_a f(T) = 0;
\end{eqnarray}
In a Friedmann-Lema\^itre-Robertson-Walker (FLRW) spatially flat spacetime, the torsion scalar takes the form
\begin{equation}
T = -6\left(\frac{\dot{a}}{a}\right)^2 \;,
\label{vincolo T}
\end{equation}
and, unlike the cosmological expression of the Ricci scalar, it does not contain second derivatives. This permits to write the cosmological point-like Lagrangian of extended TEGR without integrating higher--order terms.
\section{$f(T)$ Cosmology}
\label{f(T)}
Let us  start by applying the Noether Symmetry Approach to $f(T)$ cosmology, the simplest extension of TEGR. The corresponding action is given by Eq. \eqref{teleparallel action}. In the FLRW universe, tetrad fields can be chosen as:
\begin{equation}
h^a_\mu = \left(\begin{matrix}
1 & 0 & 0 & 0 \\ 0 & a & 0 & 0 \\ 0 & 0 & a & 0 \\ 0 & 0 & 0 & a 
\end{matrix}\right) \;\:\;\;\;\;\; \to \;\;\; h = a^3 \;,
\label{matrix}
\end{equation}
so that the correspondence \eqref{deftet} is respected. Notice that the diagonal set of tetrads \eqref{matrix} is not the only one leading to a cosmological spatially flat line element. In fact, while in GR the metric is uniquely determined once given the interval, in the teleparallel context, different tetrad fields can yield the same line element. 

In order to find a suitable point-like Lagrangian, we use the Lagrange multipliers method with the constraint given by Eq. \eqref{vincolo T}, \emph{i.e.} $\displaystyle T = -6H^2$, with $H$ being the Hubble parameter. In this way, all  dynamical variables  depend only on the cosmic time and the three-dimensional surface term can be easily integrated. The action therefore reads: 
\begin{equation}
S = 2 \pi^2 \int \{a^3 f(T) - \lambda (T+6H^2)\} dt \;.
\end{equation}
By varying the action with respect to the torsion, we get the form of the Lagrange Multiplier $\lambda$
\begin{equation}
\lambda = a^3 \frac{\partial f(T)}{\partial T}
\end{equation}
and the point-like Lagrangian turns out to be
\begin{equation}
{\cal L} = a^3 [f(T) - Tf'(T)] - 6 a\dot{a}^2f'(T) \;,
\label{lagrangiana f(T)}
\end{equation}
where the prime denotes the derivative with respect to $T$. The Euler-Lagrange equations and the energy condition for the two variables $\{a,T\}$ are, respectively
\begin{equation}
\frac{d}{dt}\frac{\partial {\cal L}}{\partial \dot{a}} = \frac{\partial {\cal L}}{\partial a} \; \; \; \; \; \; \; \; \frac{d}{dt}\frac{\partial {\cal L}}{\partial \dot{T}} = \frac{\partial {\cal L}}{\partial T}
\label{ELaf(T)}
\end{equation}
\begin{equation}
EC \to  \dot{a} \frac{\partial \Lagr}{\partial \dot{a}} + \dot{T} \frac{\partial \Lagr}{\partial \dot{T}} - \Lagr = 0 \;,
\label{ECf(T)}
\end{equation}
and can be solved only after selecting the form of the function. Note that the Lagrangian is independent of $\dot{T}$, so that the second Euler Lagrange equation provides the constraint imposed on the torsion scalar:
\begin{equation}
a^2 T + 6 \dot{a}^2 = 0 \; \; \to \; \; T = -6 \left(\frac{\dot{a}}{a}\right)^2 \;.
\end{equation}
The Euler-Lagrange equation with respect to $a$ and the energy condition lead to the following system of differential equations:
\begin{equation}
\begin{cases}
\displaystyle \ddot{a} + \frac{\dot{a}^2}{2a} + \dot{a}\dot{T} \frac{f''(T)}{f'(T)} - \frac{1}{4} a \; \frac{T f'(T) - f(T)}{f'(T)} = 0
\\
\displaystyle a^2 [f(T) - T f'(T)] + 6 \dot{a}^2 f'(T) = 0 \;.
\end{cases}
\label{EL f(T)}
\end{equation}
Let us now apply the approach outlined in Appendix \ref{Noeth} to the Lagrangian \eqref{lagrangiana f(T)}.
It is worth noticing that the configuration space is ${\cal Q}\equiv\{a,T\}$ and the related tangent space is ${\cal TQ}\equiv\{a,\dot{a},T,\dot{T}\}$. According to the discussion in the Introduction, $\cal Q$ is the minisuperspace on which we can develop our Quantum Cosmology.

\subsection{Noether symmetries in $f(T)$ cosmology}
Following Appendix \ref{Noeth}, the  Noether vector in the minisuperspace ${\cal Q}\equiv\{a,T\}$ reads as
\begin{equation}
X = \alpha \partial_a + \beta \partial_T + \dot{\alpha}\partial_{\dot{a}} + \dot{\beta}\partial_{\dot{T}}
\end{equation}
with $\alpha = \alpha(a,T)$, $\beta = \beta(a,T)$ being the components of the infinitesimal generator $\eta^i$. 

In order to select symmetries, we impose the vanishing Lie derivative of the Lagrangian \eqref{lagrangiana f(T)}, 
that is 
\begin{equation}
 L_X{\cal L}=0\,. 
 \end{equation}
 This procedure yields the following system of three partial differential equations:
\begin{equation}
   \begin{cases}
\alpha f'(T) + 2a f'(T) (\partial_a \alpha) + \beta a f''(T) = 0 
\\
\partial_T \alpha = 0 
\\
 3\alpha a^2 f(T) - 3\alpha a^2 T f'(T) - \beta a^3 T f''(T) = 0    \;.
\end{cases}
\label{sistema f(T)}
\end{equation}
After some simple manipulations, from the above system it is possible to get the infinitesimal generators $\alpha$ and $\beta$ as well as the form of the functions:
\begin{equation}
\alpha = \alpha_0 a^{1 - \frac{3}{2n}} \;\;\;\;\;\;\;\;\;\;\;\;\;\;\; \beta = \frac{-3\alpha_0}{n}T a^{- \frac{3}{2n}}, \label{alpha e beta}
\end{equation}
\begin{equation}
f(T) =f_0 T^n \;.
\label{f(T) = T^n}
\end{equation}
Therefore,  $f(T) =f_0 T^n$ is the only function admitting  Noether symmetries \cite{Spyros}. Replacing into the equations of motion \eqref{EL f(T)}, we have 
\begin{equation}
f(T) = f_0 \sqrt{T} \;,
\end{equation}
in agreement with the results provided in Ref. \cite{Skugoreva:2017vde}. This means that the equations of motion further constrain the free parameter $n$ to $n = 1/2$. Notice that a linear combination of the Euler-Lagrange equations gives the vacuum field equations
\begin{equation}
\begin{cases}
12 \dot{a}^2 f_T(T) + a^2 f(T) = 0
\\
12 \dot{a}^2 f_{TT}(T) - a^2 f_T(T) = 0.
\end{cases}
\end{equation}
Considering the cosmological expression of  torsion, the above equations become
\begin{equation}
\begin{cases}
2 T f_T(T) - f(T) = 0
\\
2 f_{TT}(T) + f_T(T) = 0,
\end{cases}
\end{equation}
which are identically satisfied for $f(T) = \sqrt{T}$. 

Let us now discuss the Hamiltonian formalism and the possible applications to Quantum Cosmology.  Finding the generators $\alpha$ and $\beta$, with the help of eq. \eqref{relzioni per cambio variabili}, we can pass from the minisuperspace of the initial variable to the reduced space containing a cyclic variable. This will allow an exact integration of the field equations.
\subsection{The Hamiltonian formalism and the Wave Function of the Universe}  \label{HWF}

Relations \eqref{relzioni per cambio variabili} allow to introduce a cyclic variable into the system, passing from the minisperspace ${\cal{Q}}= \{ a,T\}$ to ${\cal{Q}}' = \{z,w\}$  with $z$ being a cyclic variable, that is:
\begin{equation}
\begin{cases}
\displaystyle \alpha \partial_a z + \beta \partial_T z = 1
\\
 \displaystyle \alpha \partial_a w + \beta \partial_T w = 0 \;.
\end{cases}
\label{cambio variabili}
\end{equation}
By replacing the expressions of $\beta$ and $\alpha$ in Eq. \eqref{cambio variabili}, a possible solution of the above system is
\begin{equation}
\begin{cases}
\displaystyle w = a^3 T^n
\\
 \displaystyle z = \frac{2n}{3\alpha_0}a^{\frac{3}{2n}} \;,
\end{cases}
\label{cambiov}
\end{equation}
\\
or, equivalently
\begin{equation}
\begin{cases}
\displaystyle a = \left(\frac{3\alpha_0 z}{2n} \right)^{\frac{2n}{3}}
\\
\displaystyle T = w^{\frac{1}{n}} \left(\frac{3\alpha_0 z}{2n} \right)^{-2},
\end{cases}
\end{equation}
so that the  Lagrangian written in terms of the new variables reads
\begin{equation}
{\cal L} = w(1-n) -6n \alpha_0^2 \dot{z}^2 w^{\frac{n-1}{n}} \;,
\label{new lagr f(T)}
\end{equation}
\\
where $z$ is  a cyclic variable. It is worth noticing that, once changed the minisuperspace coordinates, the equations of motion coming from  Lagrangian \eqref{new lagr f(T)} are  simpler than those in Eq. \eqref{ELaf(T)}; they take the form:
\begin{eqnarray}
&& z \to n w \ddot{z} + \dot{w} \dot{z} (n-1) = 0 \label{ELz}
\\
&& w \to w = - 6^n  \alpha_0^{2n} \dot{z}^{2n}.
\label{ELw}
\end{eqnarray} 
From Lagrangian \eqref{new lagr f(T)}, with a straightforward Legendre transformation, we  get the Hamiltonian:
\begin{equation}
\displaystyle {\cal H} = w(n-1) + \frac{3 \pi^2_z}{24 n \alpha^2_0 w^{\frac{n-1}{n}}} \;.
\label{Hamf(T)}
\end{equation}
The canonical quantization procedure can be pursued by replacing the operator $i \partial_z$ to the momentum $\pi_z$ and, by applying the relation \eqref{vincoli secondari} to the Hamiltonian \eqref{Hamf(T)},  we get the  system
\begin{equation}
\begin{cases}
\displaystyle {\cal H} \psi = 0
\\
\displaystyle i\partial_z \psi = \Sigma_0 \psi 
\end{cases}
\end{equation}
where  the first is  Wheeler-DeWitt equation and the second is the conserved momentum. Explicitly, we have 
\begin{equation}
\begin{cases}
\displaystyle \left[w(n-1) - \frac{3 \partial^2_z}{24 n \alpha^2_0 w^{\frac{n-1}{n}}}\right]\psi = 0
\\
\displaystyle i\partial_z \psi = \Sigma_0 \psi  \;.
\end{cases}
\end{equation}
The solution of the second equation is
\begin{equation}
\psi = \psi_0 e^{i\Sigma_0 z} \;,
\end{equation}
which, combined with the first, yields the Wave Function of the Universe, that is:
\begin{equation}
\psi \sim \exp \left\{i\left[2 \alpha_0 w^{\frac{2n -1}{2n}} \sqrt{2n(n-1)}\right]z\right\} \;.
\end{equation}
This wave function is oscillating so, according to  the Hartle criterion, permits to select classical trajectories (see  \cite{Capozziello:1999xr} for a detailed discussion). By recasting $\psi$ in terms of the action $S_0$ as
\begin{equation}
\psi \sim \exp\left\{i S_0 \right\} \;,
\end{equation}
and identifying the action with the quantity
\begin{equation}
\nonumber S_0 = \left[2 \alpha_0 w^{\frac{2n -1}{2n}} \sqrt{2n(n-1)}\right]z,
\end{equation}
the Hamilton-Jacobi equations give:
\begin{equation}
\begin{cases}
\displaystyle \frac{\partial S_0}{\partial z} = \pi_z = \Sigma_0
\\
\\
\displaystyle \frac{\partial S_0}{\partial w} = \pi_w = 0 \;.
\end{cases}
\label{ELWF}
\end{equation} 
Note that the above equations are nothing but the Euler-Lagrange equations for the new variables defined in Eqs. \eqref{ELz} and \eqref{ELw}. The general solution is
\begin{equation}
z(t) = z_0 t \quad w =  - 6^n  \alpha_0^{2n} z_0^{2n}.
\end{equation}
Coming back to the old variables, the scale factor and the torsion scalar can be written as functions of time 
\begin{equation}
\label{classical}
a(t) = a_0 t^{\frac{2n}{3}} \; \;\;\; \; T(t) = -\frac{8 n^2}{3} \frac{1}{t^2} 
\end{equation}
for any $f(T) = T^n$ except for $n = 1/2$. Once merged with the energy condition $\left(\dot{z} \partial_{\dot{z}} + \dot{w} \partial_{\dot{w}} - 1 \right) \Lagr = 0$, the equations are further constrained to the subcase $n=1/2$, which turns out to be a trivial case, as pointed out at the beginning of this section.  In summary, the existence of the Noether symmetry allowed to integrate exactly the dynamical system and, furthermore, it allows to find out classical trajectories in the minisuperspace which can be interpreted as observable universe.  In the specific case of \eqref{classical}, depending on the value of $n$, we can achieve Friedmann decelerating solutions (for $0<n<3/2)$  or power-law inflationary solutions.
\section{$F(T,\Box T)$ Cosmology}
\label{HO}
A further extension of TEGR can come considering higher-order derivative terms in the torsion scalar as  $\Box T$, being $\Box$ the d'Alembert operator $\Box = D_\mu D^\mu$ \cite{Capriolo2}. It represents a case of particular interest, since the corresponding metric theory $F(R,\Box R)$  is renormalizable at one-loop and higher-loop level \cite{Gottlober:1989ww, Berkin:1990nu, Capozziello:1999xs}. In general, higher-order terms  arise by considering  quantum corrections to GR,  as shown in \cite{Birrell:1982ix, Amendola:1993bg} and then it is worth studying their effects also in TEGR.

In principle, the expression of the operator $\Box$ is different from  the  one in GR, due to the different form of  connections. In this case, the Weitzenb\"ock connection introduced in Sec. \ref{TG} yields: 
\begin{eqnarray}
\Box T &=& \nabla_\mu \nabla^\mu T = \nabla_\mu \partial^\mu T = \partial_\mu \partial^\mu T + \Gamma^p_{\mu p} \partial^\mu T = \nonumber
\\
&=&\ddot{T} + \Gamma^p_{0 p} \dot{T} = \ddot{T} + 3\left( \frac{\dot{a}}{a}\right) \dot{T} \;.
\label{relazbox}
\end{eqnarray}
Note that despite the different form of the connection, the d'Alembert operator has the same expression as in GR.

Thanks to Eq. \eqref{relazbox}, we can use the Lagrange multipliers method in order to get the point-like Lagrangian; to develop the approach, we treat the torsion $T$ and its d'Alembertian $\Box T$ as separated fields, although they are related through Eq.\eqref{relazbox}. Let us start by writing the general action
\begin{equation}
S = \int h F(T,\Box T) \; d^4x \;
\end{equation} 
so that, after integrating the three-dimensional hyper-surface, considering the  Lagrange multiplier gives:
\begin{eqnarray}
S &=& 2 \pi^2 \int \left\{a^3 F(T,\Box T) - \lambda_1\left(T + 6\frac{\dot{a}^2}{a^2} \right) + \right. \nonumber
\\
&-& \left. \lambda_2 \left( \Box T - \ddot{T} - 3 \frac{\dot{a}}{a} \dot{T} \right) \right\} dt \;.
\label{telepmiddleact}
\end{eqnarray}
with $\lambda_1$ and $\lambda_2$ being the two Lagrange multipliers. By varying the action with respect to $T$ and $\Box T$ we find:
\begin{equation}
\lambda_1 = a^3 \frac{\partial F(T,\Box T) }{\partial T} 
\label{lambda1}
\end{equation}
and
\begin{equation}
 \lambda_2 = a^3 \frac{\partial F(T,\Box T) }{\partial \Box T} \;.
 \label{lambda2}
\end{equation}
Replacing Eqs.\eqref{lambda1} and \eqref{lambda2} into the action \eqref{telepmiddleact} and integrating the second-order time derivatives, we finally get 
\begin{eqnarray}
\Lagr &=& a^3 \left(F - T \frac{\partial F}{\partial T} - \Box T \frac{\partial F}{\partial \Box T} \right) + \nonumber
\\
&-& 6 a \dot{a}^2 \frac{\partial F}{\partial T} - a^3 \dot{T} \dot{\Box T} \frac{\partial^2 F}{\partial \Box T^2} - a^3 \dot{T}^2 \frac{\partial^2 F}{\partial T \partial \Box T} \;.
\label{lagrangiana f(t box t)}
\end{eqnarray}
Note that by setting $\displaystyle \frac{\partial F}{\partial \Box T} = 0$ we recover the Lagrangian \eqref{lagrangiana f(T)}, discussed in Sec. \ref{f(T)}. The Euler-Lagrange equation with respect to $\Box T$ provides the cosmological expressions of $T$ and $\Box T$, while the Euler-Lagrange equation related to the torsion gives the constraint $\Box f_{\Box T} = 0$. The equation with respect to the scale factor, together  with the energy condition, provide the dynamics of the variables in the minisuperspace ${\cal Q}\equiv \{a,T,\Box T\}$ whose tangent space is ${\cal T Q}\equiv \{a,\dot{a}, T,\dot{T}, \Box T, \dot{\Box T}\}$. 
 The complete system of differential equations reads as
\begin{eqnarray}
&&4 a \ddot{a} F_T(T,\Box T) + 4 a \dot{a} \dot{\Box T} F_{T \Box T} (T,\Box T) +  \nonumber
   \\
   &&+4 a \dot{a} \dot{T} F_{TT}(T,\Box T) + 2 \dot{a}^2 F_T(T,\Box T) + \nonumber
   \\
   &&-  a^2 \dot{\Box T} \dot{T}
   F_{\Box T \Box T} (T,\Box T) - a^2 \dot{T}^2 F_{T \Box T}(T,\Box T) + \nonumber
   \\
   &&-  a^2 \Box T 
   F_{\Box T}(T,\Box T) -  a^2 T F_T (T,\Box T) + \nonumber
   \\
   &&+ a^2 F(T,\Box T) = 0   \nonumber
   \\ \nonumber
   \\
   && \Box f_{\Box T}  = 0 \nonumber
   \\ \nonumber
   \\
   && \frac{\partial^2  F(T, \Box T)}{\partial \Box T^2} \left(\Box T - 3 \frac{\dot{a}}{a} \dot{T} - \ddot{T} \right) +\nonumber
\\
&+& \frac{\partial^2  F(T, \Box T)}{\partial T \partial \Box T} \left(T + 6 \frac{\dot{a}^2}{a^2} \right) = 0  \nonumber
\\ \nonumber
\\
&& 6 \dot{a}^2 F_T(T, \Box T) + a^2 \dot{\Box T} \dot{T} F_{\Box T \Box T}(T, \Box T) + \nonumber
\\
&&+ a^2 \dot{T}^2 F_{T \Box T}(T, \Box T) - a^2 \Box T F_{\Box T}(T,\Box T) +  \nonumber
\\
&& - a^2 T F_T(T, \Box T) + a^2 F(T, \Box T) = 0 \;.
\end{eqnarray}
As before, the above equations cannot be solved until the function $F(T,\Box T)$ is selected.
\subsection{Noether symmetries for  $F(T, \Box T)$ cosmology}
The application of the condition $L_X \Lagr = 0$ to the Lagrangian \eqref{lagrangiana f(t box t)} allows to find the symmetries of the high-order theory $F(T, \Box T)$. The minisuperspace of configurations contains three variables, so that Noether's vector has the form:
\begin{equation}
X = \alpha \partial_a + \beta \partial_T + \gamma \partial_{\Box T} + \dot{\alpha} \partial_{\dot{a}} + \dot{\beta} \partial_{\dot{T}} + \dot{\gamma} \partial_{\dot{\Box T}}.
\end{equation}
With respect to the previous section, the theory contains a new infinitesimal generator $\gamma$, related to the presence of the new variable $\Box T$. Imposing the condition $X \Lagr = 0$ and equating  to zero the coefficients of $\dot{a}^2 , \dot{T} \dot{\Box T}, \dot{T}^2, \dot{a} \dot{T}, \dot{a} \dot{\Box T}$ and $\dot{ \Box T}^2$, we get a system of seven differential equations:
\begin{equation}
\begin{split}
& 3 \alpha \left( F - T \frac{\partial F}{\partial T} - \Box T \frac{\partial F}{\partial \Box T} \right) - \beta a T \frac{\partial^2 F}{\partial T^2} +
\\
&- \beta a \Box T \frac{\partial^2 F}{\partial T \partial \Box T} - \gamma a T \frac{\partial^2 F}{\partial T \partial \Box T} - \gamma a \Box T \frac{\partial^2 F}{\partial \Box T^2} = 0
\\
& \left(\alpha + 2a \frac{\partial \alpha}{\partial a} \right) \frac{\partial F}{\partial T} + \beta a \frac{\partial^2 F}{\partial T^2} + 
\\
&+\gamma a \frac{\partial^2 F}{\partial T \partial \Box T} = 0
\\
& \left(3 \alpha + a \frac{\partial \gamma}{\partial \Box T} + a \frac{\partial \beta}{\partial T} \right) \frac{\partial^2 F}{\partial \Box T^2} + \beta a \frac{\partial^3 F}{\partial T \partial \Box T^2} +
\\
&+ \gamma a \frac{\partial^3 F}{\partial \Box T^3} = 0
\\
& \left(3 \alpha + 2a \frac{\partial \beta}{\partial T} \right) \frac{\partial^2 F}{\partial T \partial \Box T} + \beta a \frac{\partial^3 F}{\partial T^2 \partial \Box T} + 
\\
&+\gamma a \frac{\partial^3 F}{\partial T \partial \Box T^2} + a \frac{\partial \gamma}{\partial T} \frac{\partial^2 F}{\partial \Box T^2} = 0
\\
&12 \frac{\partial \alpha}{\partial T} \frac{\partial F}{\partial T} + a^2 \frac{\partial \gamma}{\partial a} \frac{\partial^2 F}{\partial \Box T^2} + 2a^2 \frac{\partial \beta}{\partial a} \frac{\partial^2 F}{\partial T \partial \Box T} = 0
\\
& 12 \frac{\partial \alpha}{\partial T} \frac{\partial F}{\partial T} + a^2 \frac{\partial \beta}{\partial a} \frac{\partial^2 F}{\partial \Box T^2} = 0
\\
& \frac{\partial \beta}{\partial \Box T} = 0 \;,
\end{split}
\label{sistema 7eq}
\end{equation} 
whose non trivial solutions are
\begin{equation}
\begin{cases}
\displaystyle \alpha = 0
\\
\displaystyle \beta = \beta_0
\\
\displaystyle \gamma = \beta_0 \frac{\Box T}{T}
\\
\displaystyle F_{I}(T, \Box T) = F_0 T^{1-k} \Box T^k
\end{cases}
\begin{cases}
\displaystyle \alpha = - \frac{1}{3} \gamma_0 k \Box T a^{-\frac{1}{2}}
\\
\displaystyle \beta = \beta_0 a^{-\frac{3}{2}}
\\
\displaystyle \gamma = \gamma_0 (\Box T)^2 a^{-\frac{3}{2}}
\\
\displaystyle F_{II}(T, \Box T) =  f_0 (\Box T)^k +
\\+ f_1 T
\end{cases} \nonumber
\end{equation}
\begin{equation}
\begin{cases}
\displaystyle \alpha = - \frac{1}{3} \gamma_0 k a \Box T \chi(a)
\\
\displaystyle \beta =  \beta_0 \chi(a)
\\
\displaystyle \gamma =  \gamma_0 (\Box T)^2 
\\
\displaystyle F_{III}(T, \Box T) = f_0 (\Box T)^k 
\end{cases}
\begin{cases}
\displaystyle \alpha = \alpha_0 a^{1 - \frac{3}{2k}}
\\
\displaystyle \beta = - \frac{3 \alpha_0}{k} T a^{- \frac{3}{2k}} 
\\
\displaystyle \gamma = \gamma(a,T,\Box T)
\\
\displaystyle F_{IV}(T, \Box T) =  f_0 T^k + f_1 \Box T.
\end{cases}
\label{solutioni f t box t}
\end{equation} 
The first function is the only one made of a product of $T$ and $\Box T$; it is of particular interest, since the product between the two variables can be seen as a modified $f(T)$ theory of gravity non-minimally coupled to a scalar field. The infinitesimal generators of the second solution contain a  function of the scale factor; the corresponding function consists in a sum of the two variables $T$ and $\Box T$, as well as the fourth function. Specifically, this latter leads to the same Lagrangian as $f(T)$ gravity in Eq. \eqref{lagrangiana f(T)}. This means that the additive contribution of the higher-order term $\Box T$ does not introduce any further degrees of freedom; in such a case, also the generator coefficients are, in turn, the same as those found in Sec. \ref{f(T)}. 

In what follows,  we focus on the first solution of Noether system, discussing the Hamiltonian dynamics and solving the Wheeler-DeWitt equations in the minisuperspace of the three variables $a,T,\Box T$. The point-like Lagrangian can be found by replacing the function $F_I(T, \Box T)$ into Eq. \eqref{lagrangiana f(t box t)}, obtaining
\begin{eqnarray}
\label{lagrangiana finale f(t box t}
\Lagr &=& F_0(k-1) \left\{ 6 a \dot{a}^2 \left( \frac{\Box T}{T}\right)^k - ka^3 \dot{T} \dot{\Box T} \left( \frac{\Box T}{T}\right)^{k-2} \frac{1}{T} + \right. \nonumber
\\
&+& \left. k a^3 \dot{T}^2 \left( \frac{\Box T}{T}\right)^{k-1}  \frac{1}{T} \right\} \;.
\end{eqnarray}
The Euler-Lagrange equations can be analytically solved for $k \ge 2$ providing a de Sitter-like cosmological solution of the form
\begin{eqnarray}
a(t) &=& a_0 e^{nt}, \;\;\; T(t) = -6n^2, \;\;\; \Box T(t) = 0,
\label{soluzioni cosmologiche t box t}
\end{eqnarray}
with $n$ being an arbitrary real number. To find the Hamiltonian and the corresponding wave function, we use the condition in Eq. \eqref{relzioni per cambio variabili}, by means of which the minisuperspace ${\cal{Q}} \equiv \{a,T, \Box T \} $ can be transformed to $ {\cal{Q}}'\equiv \{z, \omega, u\}$, where $z$ is a cyclic variable. The change of variables yields the system
\begin{equation}
\begin{cases}
\displaystyle \alpha \frac{\partial z}{\partial a} + \beta \frac{\partial z}{\partial T} + \gamma \frac{\partial z}{\partial \Box T} = 1
\\
\\
\displaystyle \alpha \frac{\partial \omega}{\partial a} + \beta \frac{\partial \omega}{\partial T} + \gamma \frac{\partial \omega}{\partial \Box T} = 0
\\
\\
\displaystyle \alpha \frac{\partial u}{\partial a} + \beta \frac{\partial u}{\partial T} + \gamma \frac{\partial u}{\partial \Box T} = 0 \;.
\end{cases}
\label{cambio variabili generico}
\end{equation}
whose possible solution is
\begin{equation}
z = z(T) = \frac{T}{\beta_0}, \;\;\;\;\; \omega = \frac{\Box T}{T} ,\;\;\;\;\; u = a \;,
\end{equation}
from which the variables $T$ and $\Box T$ can be written in terms of the new variables $\omega$ and $z$ as:
\begin{equation}
T = \beta_0 z, \;\;\;\;\;\;\; \Box T = \beta_0 \omega z, \;\;\; a=u
\label{vecchie in funzione delle nuove}
\end{equation}
The Lagrangian $\Lagr(z, \omega, u)$ turns out to be\footnote{Since $u=a$ we prefer to write Lagrangian depending on $a$ instead of $u$.}
\begin{equation}
 \Lagr_{new} = F_0(k-1) \left\{6 a \dot{a}^2 \omega^k - k \beta_0 a^3 \omega^{k-2} \dot{z} \dot{\omega} \right\} \;.
\label{lagrangiana f(t box t) nuova}
\end{equation}
As expected, the shape of the Lagrangian suggests that $z$ is now a cyclic variable, so that the constant of motion $\Sigma_0$ is the conjugate momentum related to the cyclic variable: 
\begin{equation}
\Sigma_0 = \frac{\partial \Lagr}{\partial \dot{z}} = - F_0 \beta_0 k(k-1) a^3 \omega^{k-2} \dot{\omega} \;.
\end{equation}
\subsection{The Hamiltonian and Wave Function of the Universe}
For the sake of simplicity, we define $F_0 k(k-1) \beta_0 = p_0$ and $12F_0(k-1) = p_1$, so that the conjugate momenta can be written as:
\begin{equation}
\label{momenti 3 variabili}
\begin{cases}
\displaystyle \pi_z = \frac{\partial \Lagr}{\partial \dot{z}} = -p_0 a^3 \omega^{k-2} \dot{\omega} 
\\
\displaystyle \pi_\omega = - p_0 a^3 \omega^{k-2} \dot{z}
\\
\displaystyle \pi_a = p_1 a \omega^k \dot{a} 
\end{cases}
\end{equation}
and, finally, the Hamiltonian reads
\begin{equation}
{\cal{H}} = - \frac{1}{p_0 a^3 \omega^{k-2}} \left(\pi_z^2 + \pi_\omega^2 - \pi_z \pi_\omega \right) - \frac{1}{2} \frac{\pi_a^2}{p_1 a \omega^k} \;.
\end{equation}
The constraints imposed by Eq. \eqref{vincoli secondari sui momenti} provide the system of differential equations
\begin{equation}
\begin{cases}
\displaystyle \left(\partial_z^2 + \partial_\omega^2 + \partial_z \partial_\omega \right) \psi + \frac{1}{2} \frac{p_0}{p_1} \frac{a^2}{\omega^2} \partial_a^2 \psi = 0
\\
\displaystyle -i \partial_z \psi = \Sigma_0 \psi \;.
\end{cases}
\end{equation}
The solution of the former equation is a linear combination of Bessel functions, whose asymptotic limit provides an oscillating wave function of the form
\begin{eqnarray}
\psi \sim && \exp\left\{i\left[\frac{1}{2} \ln \omega - \frac{\sqrt{3}}{2} \Sigma_0 \omega - \frac{\pi}{4}\left(1 + \sqrt{4c_0 + 1} \right) + \right. \right. \nonumber
\\
&-& \left. \left. \Sigma_0 z + \sqrt{c_0} \ln a \right]  \right\} 
\label{psi finale}
\end{eqnarray}
Notice that the Hartle criterion is then recovered for large scale factors, where the wave function is peaked in the minisuperspace $\cal{Q}'$.
Moreover, classical trajectories can be recovered by identifying the exponential factor of Eq. \eqref{psi finale} with $S_0$, namely:
\begin{equation}
S_0 = \frac{1}{2} \ln \omega - \frac{\sqrt{3}}{2} \Sigma_0 \omega - \frac{\pi}{4}\left(1 + \sqrt{4c_0 + 1} \right) - \Sigma_0 z + \sqrt{c_0} \ln a
\label{S00}
\end{equation}
and the wave function can be recast as $\psi \sim e^{i S_0}$. The Hamilton-Jacobi equations coming from the action \eqref{S00} read as
\begin{equation}
\begin{cases}
\displaystyle \frac{\partial S_0}{\partial a} = \pi_a
\\
\\
\displaystyle \frac{\partial S_0}{\partial z} = \pi_z
\\
\\
\displaystyle \frac{\partial S_0}{\partial \omega} = \pi_\omega \;,
\end{cases}
\end{equation}
with
\begin{equation}
\pi_a = \frac{\sqrt{c_0}}{a} \;\;\;\;\;\; \pi_\omega = \frac{1}{2 \omega} -\frac{\sqrt{3}}{2} \Sigma_0  \;\;\;\;\;\; \pi_z = \Sigma_0 \;,
\end{equation}
from which a system of three differential equations follows:
\begin{equation}
\begin{cases}
\displaystyle  \; p_1 a \omega^k \dot{a} = \frac{\sqrt{c_0}}{a}
\\
\displaystyle \;  p_0 a^3 \omega^{k-2} \dot{z} = \frac{\sqrt{3}}{2} \Sigma_0 - \frac{1}{2 \omega}
\\
\displaystyle  \; p_0 a^3 \omega^{k-2} \dot{\omega} = \Sigma_0 \;.
\end{cases}.
\end{equation}
The solutions of the above system are the same as Eq. \eqref{soluzioni cosmologiche t box t}, so that  classical trajectories, and then observable universes, are immediately recovered.
\section{Teleparallel Gauss-Bonnet Cosmology}
\label{TGa}
Let us consider now the contribution of the  Gauss-Bonnet topological invariant among the possible   TEGR extensions. In  four-dimensional metric formalism, the Gauss-Bonnet scalar is a topological invariant which assumes the form:
\begin{equation}
\G = R^2 - 4R^{\mu \nu} R_{\mu \nu} + R^{\mu \nu p \sigma} R_{\mu \nu p \sigma} \;.
\label{GB}
\end{equation}
In FLRW cosmology,  the square root of the Gauss-Bonnet scalar is dynamically equivalent to the Ricci scalar, since the contribution of other second order curvature invariants are comparable with respect to that of $R^2$.  Being a topological surface term, in four dimensions, the action containing $R+\G$ provides the same dynamics as the Einstein-Hilbert action; however, in five dimensions or more the integral of $\G$ is not a topological invariant, so that the action $S = \int \left(R + \G \right) d^nx$ (with $ \ge 5$) does not yield the same equations of motion as the $n$-dimensional GR. Another issue which may be solved by introducing $\G$ in the theory, is linked to the treatment of gravity under the gauge formalism, since the Gauss-Bonnet invariant  naturally emerges in many gauge theories as Chern-Simons or Lovelock gravity. The Gauss-Bonnet term is considered as a part of the gravitational action in several works, such as \cite{Shamir:2018eru, Odintsov:2018nch} where a function of $R$ and $\G$ is discussed, or \cite{Bajardi:2020osh, Bajardi:2019zzs} where  GR is recovered from  $f(\G)$ gravity without imposing the Einstein-Hilbert action in $R$ \emph{a priori}.

Moreover, it is possible to construct a teleparallel equivalent  Gauss-Bonnet term as shown in \cite{Sharif:2018sgg, Zubair:2015yma, Capozziello:2016eaz}. 
 
The aim of this section is to apply the formalism of Quantum Cosmology to a function of the torsion and of the teleparallel Gauss-Bonnet invariant, finding the Wave Function of the Universe and recovering  classical trajectories related to observable universes. 

The cosmological expression of the teleparallel Gauss-Bonnet term is equivalent to that provided by the corresponding metric theory \cite{Manos}, namely:
\begin{equation}
\G = T_\G =  24\frac{\ddot{a} \dot{a}^2}{a^3} \;,
\end{equation}
from which it is easy to verify that it represents a total derivative. Considering that a function of $T_\G$ is not-trivial even in a four-dimensional spacetime, we take into account an action of the form 
\begin{equation}
S = \int |e|  \, f\left( T, T_\G \right) d^4x,
\end{equation}
so that TEGR is recovered as soon as $f(T, T_\G)= T$. By means of Noether's approach, it is possible to find out the cosmological solutions for a generic function of the teleparallel surface term $T_\G$ containing symmetries; this approach has been performed in \cite{Capozziello:2016eaz}, therefore in this section we only outline the main results. After this, we consider the ADM formalism and find the related  Wave Function of the Universe, as we did before. In the minisuperspace ${\cal Q}\equiv\{a,T,T_\G\}$, the symmetry generator is
\begin{equation}
X = \alpha(a,T,T_\G) \partial_a + \beta(a,T,T_\G) \partial_T + \gamma(a,T,T_\G) \partial_{ T_\G}.
\end{equation}
The application of the condition $L_X \Lagr = 0$ to the point-like Lagrangian
\begin{eqnarray}
\Lagr &=& a^{3}\left(f- T_\G f_{T_\G} - T f_{T}\right) \nonumber
\\
&-& 8 \dot{a}^{3}\left(T_\G f_{T_\G T_\G}+ \dot{T} f_{T T_\G}\right) -6 f_{T} a \dot{a}^{2},
\label{lagr f(T,TG)}
\end{eqnarray}
gives rise to the following Noether solutions \cite{Capozziello:2016eaz}:
\begin{equation}
X = \left\{0;\beta(a,T_\G,T); \frac{T}{T_\G} \beta(a,T_\G,T)\right\}, \;\;\; f(T,T_\G) = f_0 T_\G^k T^{1-k}
\end{equation}
and then the point-like Lagrangian reduces to 
\begin{equation}
\Lagr = f_0(k-1) \dot{a}^2 T_\G^{k-2} T^{-k} \left[4k \dot{a}(T_\G \dot{T} - T \dot{T_\G}) + 3 a T_\G^2\right]\,.
\label{Lagrangiana Gauss-Bonnet}
\end{equation}
The solutions of the  Euler-Lagrange equations are
\begin{equation}
a(t) = a_0 t^{2k+1}, \;\; T = -6 \frac{(2k+1)^2}{t^2}, \;\;\; T_\G(t) =24 \frac{2k (2k+1)^3}{t^4} \;.
\label{soluzioni cosmologiche GB}
\end{equation}
Here $k$ is an arbitrary real number and then the above considerations work, i.e. it is possible to recover Friedmann and inflationary  cosmological solutions.

In order to find the Hamiltonian, we perform the change of variables by means of which we can introduce a cyclic variable $z$; the system \eqref{relzioni per cambio variabili} can be written as:
\begin{equation}
\begin{cases}
\displaystyle \beta \partial_{T_\G} z + \beta \frac{T}{T_\G} \partial_T z = 1
\\
\displaystyle \partial_{T_\G} w + \frac{T}{T_\G} \partial_T w = 0
\\
\displaystyle \partial_{T_\G} u + \frac{T}{T_\G} \partial_T u = 0
\end{cases}
\end{equation}
and, by assuming $\beta = \beta_0 T_\G$, one possible solution is 
\begin{equation}
z = \frac{1}{\beta_0} \ln(T_\G)  \;\;\; w = \frac{T_\G}{T} \;\;\; u = a \;,
\end{equation}
from which 
\begin{equation}
a = u \;\;\; T_\G = e^{\beta_0 z} \;\;\; T = \frac{e^{\beta_0 z}}{w} \;.
\end{equation}
Thanks to these relations, Lagrangian \eqref{Lagrangiana Gauss-Bonnet} takes the form:
\begin{equation}
\Lagr =  f_0(k-1) \dot{a}^2 w^k \left[- 4k \frac{\dot{a} \dot{w}}{w^2} + 3 a\right] \;.
\end{equation}
Notice that the cyclic variable $z$ does not appear in the Lagrangian and the minisuperspace is further restricted to the two variables $w$ and $a$. By means of the conserved quantity $\Sigma_0$
\begin{equation}
\left(\frac{- w^{2-k} \pi_w}{4f_0 k(k-1)}\right)^\frac{1}{3} = \Sigma_0,
\end{equation} 
the Hamiltonian can be written as:
\begin{equation}
\mathcal{H} = \Sigma_0 \pi_a - 3f_0 k(k-1) a w^k \Sigma_0^2 \;.
\end{equation}
From the canonical quantization rules, after writing Hamiltonian in terms of operators, the Wheeler-De Witt equation takes the form
\begin{equation}
\hat{\mathcal{H}} \psi = - i \partial_a \psi- 3f_0(k-1) a w^k \Sigma_0 \psi = 0 \;,
\end{equation}
whose solution is:
\begin{equation}
\psi = \psi_0 \exp \left\{i \left[\frac{3}{2} \Sigma_0 f_0 (k-1) w^k \right] a^2 \right\} \;.
\end{equation}
The Hartle criterion is recovered and the Hamilton-Jacobi equations provide the above  classical trajectories. 
\section{Non-minimally coupled scalar field}
\label{Scalar}
The final case we are going to analyze is  teleparallel scalar-tensor TEGR, whose  action reads as:
\begin{equation}
S = \int \left[T F(\phi) + \frac{\omega}{2} \partial_\mu \phi \partial^\mu \phi - V(\phi) \right] d^4 x
\end{equation}
being $\phi$ a scalar field, $F(\phi)$ the non-minimal coupling, and $V(\phi)$ the self-interacting potential. We can also take into account  more general actions, depending on a general function $F(T, \phi$), however, here we restrict to the case linear in $T$. The above action is the teleparallel equivalent of  the well known scalar-tensor action in the metric formalism (see  \cite{Bartolo:1999sq, Capozziello:2007iu}) discussed \emph{e.g.} in \cite{Sadjadi:2013nb, Bahamonde:2018miw}. In FLRW cosmology, where the hyper-surface term can be integrated, we can recast the action as an integral over the time where variables are only time dependent, namely
\begin{equation}
\label{Azione campo scalare}
S = 2 \pi^2 \int h\left[F(\phi) T + \frac{1}{2} \dot{\phi}^2 - V(\phi) \right] dt \;,
\end{equation}
Replacing therefore the cosmological form of the torsion into the action, the Lagrangian reads:
\begin{equation}
{\cal L} = {\cal L}(a,\dot{a},\phi,\dot{\phi}) = -6 F(\phi) a \dot{a}^2 + \left[\frac{1}{2} \dot{\phi}^2 - V(\phi) \right] a^3 \;.
\label{lagra scalare}
\end{equation}
In this case, the minisuperspace only consists of  the two variables $a$ and $\phi$, that is ${\cal Q}\equiv\{a,\phi\}$; the torsion does not appear manifestly as a variable, having been replaced by its cosmological expression. Starting from the Lagrangian \eqref{lagra scalare}, the equations of motion and the energy condition $E_\Lagr = 0$ yield the system of differential equations:
\begin{equation}
\begin{cases}
\displaystyle 2 a \dot{a} \dot{\phi} F_\phi(\phi) + 4 \dot{a}^2 F(\phi) + 4 a \ddot{a} F(\phi) + \frac{1}{2} a^2 \dot{\phi}^2- a^2 V(\phi) = 0
\\
\displaystyle 3 a \dot{a} \dot{\phi} + a^2 \ddot{\phi} + 6 \dot{a}^2 F_\phi(\phi) + a^2 V_\phi(\phi) = 0
\label{eq motion scalar}
\\
\displaystyle 6 F(\phi) \dot{a}^2 - \frac{1}{2} a^2 \dot{\phi}^2 - a^2 V(\phi) = 0 \;.
\end{cases}
\end{equation}
The system can be solved after selecting the form of the coupling and the potential by Noether's symmetries.  
\subsection{Noether symmetries}

Considering the Noether vector of the two-dimensional minisuperspace ${\cal{Q}} = \{a, \phi \}$, namely
\begin{equation}
X = \alpha \partial_a + \beta \partial_\phi + \dot{\alpha} \partial_{\dot{a}} + \dot{\beta} \partial_{\dot{\phi}}
\end{equation}
and setting the Lie derivative of the Lagrangian \eqref{lagra scalare} along  $X$ equal to zero, gives the  following system of partial differential equations:
\begin{equation}
\begin{cases}
\alpha F(\phi) + \beta a F'(\phi) + 2F(\phi) a \partial_a \alpha = 0 
\\
3 \alpha + 2a \partial_\phi \beta = 0 
\\
a^2 \partial_a \beta - 12F(\phi) \partial_\phi \alpha = 0
\\
3\alpha V(\phi) + \beta a V'(\phi) = 0  \;,
\end{cases}.
\label{sistema f(fi)}
\end{equation}
whose non trivial solutions are
\begin{equation}
\begin{cases}
\displaystyle \alpha = -\frac{2}{3} \beta_0 \sqrt{V_0} a
\\
\displaystyle \beta = \beta_0 \sqrt{V_0} \phi
\\
\displaystyle V_{I}(\phi) = V_0 \phi^2
\\
\displaystyle F_{I}(\phi) = F_0 \phi^2 \;
\end{cases}
\begin{cases}
\displaystyle \alpha = - \frac{2 \beta_0}{2 \ell + 3} a^{\ell +1} \phi^{-\frac{2 \ell}{2 \ell +3}}
\\
\displaystyle \beta =  \beta_0 a^\ell \phi^{\frac{3}{2 \ell +3}}
\\
\displaystyle V_{II}(\phi) = V_0 \phi^{\frac{6}{2 \ell + 3}}
\\
\displaystyle F_{II}(\phi) =\frac{(2 \ell + 3)^2}{48} \phi^2;
\end{cases}
\nonumber
\end{equation}
\begin{equation}
\begin{cases}
\displaystyle \alpha =  \frac{2q \beta_0}{3} \frac{e^{q \phi}}{\sqrt{a}}
\\
\displaystyle \beta = \beta_0 \frac{e^{q \phi}}{\sqrt{a^3}}
\\
\displaystyle V_{III}(\phi)  = V_0 e^{2 q \phi}
\\
\displaystyle F_{III}(\phi) =  \frac{3}{16 q^2}
\end{cases}
\begin{cases}
\displaystyle \alpha = -\frac{2}{3} a^{\frac{1}{4}} (c_2 + 2 c_3 \phi)
\\
\displaystyle \beta = a^{-\frac{3}{4}}(c_1 + c_2 \phi + c_3 \phi^2)
\\
\displaystyle V_{IV}(\phi) = V_0(c_1 + c_2 \phi + c_3 \phi^2)^2
\\
\displaystyle F_{IV}(\phi) = \frac{3}{64 c_3} (c_1 + c_2 \phi + c_3 \phi^2) 
\end{cases}
\end{equation}
Let us focus on the first one, with the aim to investigate both the classical and the quantum cosmological implications. The point-like Lagrangian, which arises once replacing the functions $F_I(\phi)$ and $V_{I}(\phi)$ into Eq. \eqref{lagra scalare}, takes the form
\begin{equation}
{\cal L} = -6 F_0 a \phi^2 \dot{a}^2 + a^3\left[\frac{1}{2}\dot{\phi^2} - V_0 \phi^2 \right],
\end{equation}
so that the conserved quantity is
\begin{equation}
\Sigma_0 = \alpha \partial_a {\cal L} + \beta \partial_\phi {\cal L} = \beta_0 \sqrt{V_0} a^2 \phi(8F_0 \phi \dot{a} + a\dot{\phi}) \;.
\label{sigma}
\end{equation}
The equations of motion \eqref{eq motion scalar}, when $F \neq 0, \, V \neq 0$, only provide exponential solutions of the form
\begin{eqnarray}
&& a(t) = a_0 e^{kt} \;\;\;\,\,\, \phi = \phi_0 e^{8F_0 k (t - t_0)} \nonumber
\\
&& V(\phi) = 2 k^2 F_0(3 - 16 F_0) \phi^2 \quad F(\phi) = F_0 \phi^2.
\end{eqnarray}
By neglecting the contribution of the potential, the Euler-Lagrange equations  provide also power-law solutions, which further constrain $F(\phi)$ to
\begin{eqnarray}
&&a(t) = a_0 t^k \;\;\; \phi(t) = \phi_0 t^{\frac{1}{2} (1-3
   k)} \nonumber
   \\
   &&     F(\phi) = \frac{(3 k-1)^2}{48 k^2} \phi^2 \;.
   \label{timpow}
\end{eqnarray}
From the above power-law expressions of the scale factor, we can distinguish the three different Friedmann eras \begin{eqnarray}
\text{Radiation Dominated Era}: && \to a(t) = a_0 t^{\frac{1}{2}}  \quad \phi(t) = \phi_0 t^{- \frac{1}{4}} \nonumber
\\
\text{Stiff Matter Dominated Era}: && \to a(t) = a_0 t^{\frac{1}{3}} \quad \phi(t) = \phi_0 \nonumber
\\ 
\text{Dust Matter Dominated Era}: && \to a(t) = a_0 t^{\frac{2}{3}} \quad \phi(t) = \phi_0 t^{- \frac{1}{2}} \nonumber
\end{eqnarray}
The evolution of the scalar field  can be summarized in the figure below:

\begin{center}
\centering
\includegraphics[width=.50\textwidth]{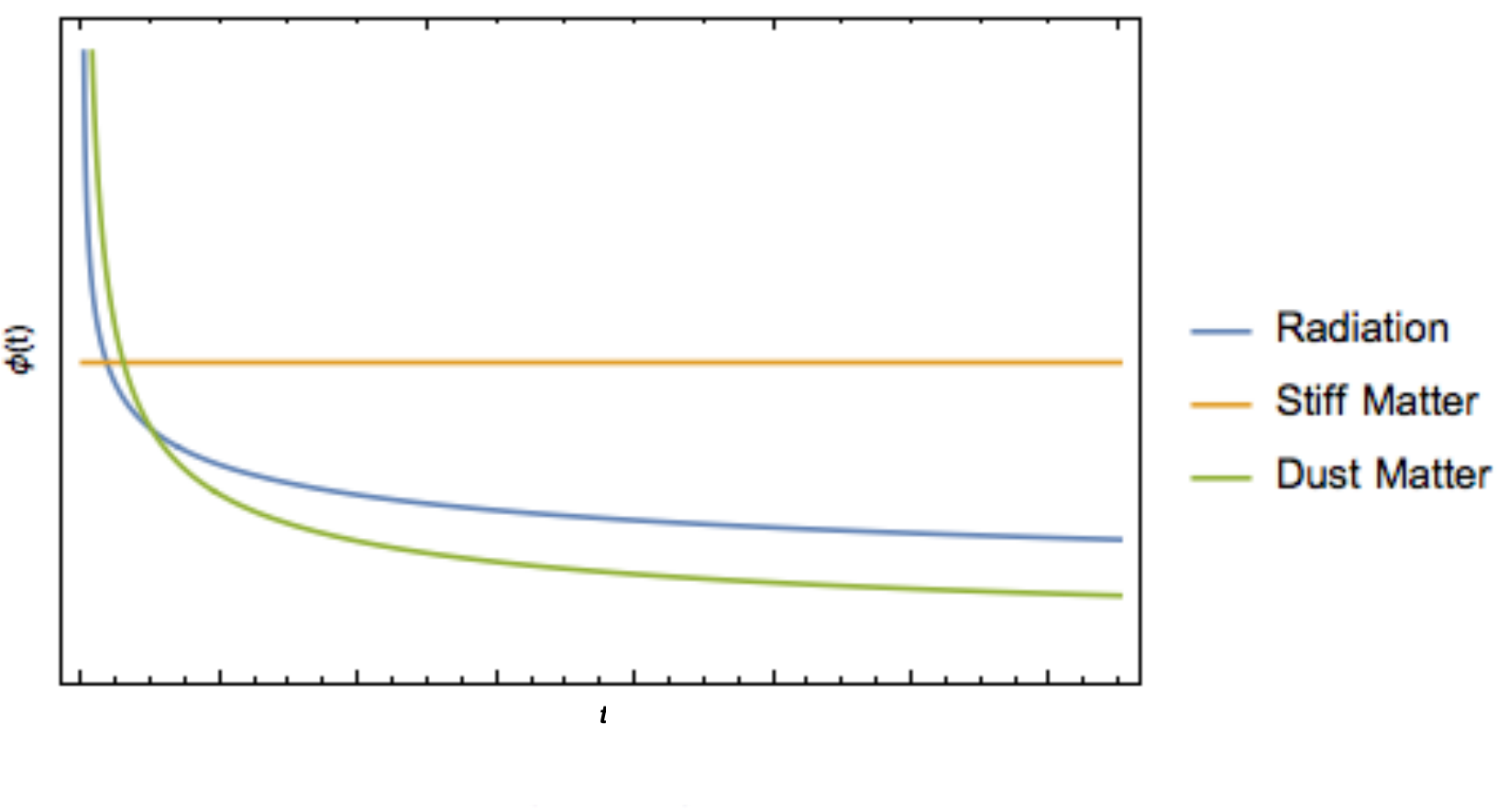}
\label{diagramma uniti}
\\ Fig. 1: \emph{Behavior of the scalar field in the three cosmological eras  as  functions of time.}
\end{center}

We now apply the relations in Eqs. \eqref{cambio variabili}  to find the cyclic variable $z$ in the minisuperspace.  It yields the system
\begin{equation}
\begin{cases}
\displaystyle -\frac{2}{3}a \partial_a z + \phi  \partial_\phi z = 1
\\
\displaystyle -\frac{2}{3} a \partial_a w + \phi \partial_\phi w = 0 \;.
\end{cases}
\end{equation}
\\
Assuming $w = a^3 \phi^2$ and $\partial_a z = 0$, one of the possible solution is
\begin{equation}
\begin{cases}
\displaystyle z = \ln \phi
\\
\displaystyle w = a^3 \phi^2
\end{cases}
\;\;\;\;\;\;\;\;\;\;\;\;
\begin{cases}
\displaystyle \phi = e^z
\\
\displaystyle a = w^{\frac{1}{3}} e^{-\frac{2}{3} z}
\end{cases}
\end{equation}
and the new Lagrangian is then
\begin{equation}
\Lagr_{new} = -\frac{2}{3} F_0 \frac{\dot{w}^2}{w} + \frac{8}{3}F_0 \dot{w}\dot{z} + f_0 w\dot{z}^2 - V_0 w \;,
\label{L new scalar field}
\end{equation}
where we set $f_0 \equiv 1 - 8/3 \; F_0$. Here $z$ is the cyclic variable and the conserved quantity can be written in terms of the momentum $\pi_z$ as:
\begin{equation}
\partial_{\dot{z}} \Lagr = \pi_z = \frac{8}{3}F_0 \dot{w} + 2 f_0 w \dot{z}= \Sigma_0.
\label{constraint}
\end{equation}
\subsection{Hamiltonian and Wave Function of the Universe}
The ADM formalism can be pursued by writing the time-derivatives of the cosmological variables as functions of momenta and performing a Legendre transformation; this procedure yields the Hamiltonian :
\begin{equation}
{\cal H} = \frac{\pi_z^2}{2w} - \frac{1}{2} \frac{f_0}{F_0} w \pi_w^2 + 2 \pi_z \pi_w + V_0 w \;.
\label{hamiltoniana}
\end{equation}
The Wave Function of the Universe is  the solution of the system 
\begin{equation}
\begin{cases}
\displaystyle -\frac{1}{2w}\partial_z^2 \psi(z,w) - 2\partial_w \partial_z \psi(z,w) + 
\\
\displaystyle +\frac{1}{2} \frac{f_0}{F_0} w \partial_w^2 \psi(z,w) + V_0 w \psi(z,w) = 0
\\
\\
\displaystyle i\partial_z \psi(z,w) = \Sigma_0 \psi(z,w) \;.
\end{cases}
\label{sepvariabili}
\end{equation}
The first equation is the Wheeler-De Witt equation, coming from  $\hat{\cal{H}} \psi = 0$; the second equation is  conservation relation provided by the Noether symmetry, namely $\pi_z = \Sigma_0$. The solution of \eqref{sepvariabili} is a linear combination of Bessel's functions, which, for large arguments,  are peaked in the minisuperspace  variables $z,w$. Therefore the wave function has an asymptotic behavior of the form
\begin{equation}
\psi(z,w) \sim e^{i\left(\Sigma_0 z - \ln w + Ww - p \frac{\pi}{2} - \frac{\pi}{4} \right)} \;,
\label{psi}
\end{equation}
where $p$, $A$, $B$ and $W$ are defined as
\begin{eqnarray}
&& p = \frac{1}{2} \sqrt{-A^2 -2iA -4B + 1} \nonumber
\\
 &&  -4 \Sigma_0 \frac{F_0}{f_0} = A; \,\, \,\,\;\; \frac{2F_0}{f_0}V_0 = W^2; \;\;\;\;\; \frac{F_0}{f_0} \Sigma_0^2 = B.
\end{eqnarray}
Notice that Hartle's criterion is recovered after imposing the change of variables suggested by the Noether symmetry; furthermore, in the semiclassical limit where the wave function can be recast as $\psi \sim e^{i S_0}$, the Hamilton-Jacobi equations yield a system of two differential equations of the form
\begin{equation}
\begin{cases}
\pi_z = \partial_z S_0 = \Sigma_0  
\\
\pi_w = \partial_w S_0 = W -\frac{1}{w} \;,
\end{cases}
\label{syst}
\end{equation}
whose solution is
\begin{equation}
a(t) \sim e^{kt} \;\;\; \phi(t) \sim e^{\ell t} \;.
\end{equation}
Eqs. \eqref{syst} provide the same exponential solution as the Euler-Lagrange equations, as expected by construction. The  power-law solution \eqref{timpow} can be recovered for the  potential $V(\phi)=0$. Notice that even in this case Noether symmetry  allows to  calculate the equations of motion solutions by means of the semiclassical limit of the ADM formalism.
\section{Discussion and Conclusions}
\label{concl}

The  purpose of this paper is to apply the Noether Symmetries Approach to different extended  TEGR models in view of Quantum Cosmology applications.  In several works \cite{Capozziello:2013qha, Capozziello:2014qla, Capozziello:2014ioa, Vakili:2008ea}, the Noether Approach has been used  to select the Lagrangians depending on functions of curvature invariants, using the symmetries and reducing therefore the dynamics. In this paper we found out exact solutions coming from actions containing functions of  torsion scalar $T$ and its related invariants. In all cases, we dealt with the quantum counterpart by using  the ADM formalism.

TEGR  models have been proposed with the aim to relax the strict constraints provided by GR, as well as the strict dependence on Equivalence  Principle, metricity or Lorentz invariance; another assumption of GR  concerns the Levi-Civita connection supposed to be torsionless. Relaxing the hypothesis of torsionless connection, it is possible to construct  spacetimes where affinities have a dynamical role, instead of geodesic structure. TEGR  is completely equivalent to GR at the level of equations and then of dynamics. This can be easily understood by the fact that TEGR  and GR field equations differ only  for a 4-divergence. However, though TEGR provides the same results as GR, extending TEGR turns out to be  different with respect to GR extensions, since this latter leads to higher-order field equations with respect to the metric. Nevertheless, $f(R)$ gravity can be always obtained from $f(T)$ gravity considering  the boundary term $B$: a function $f(T,B) $, under appropriate constraints, allows to recover the $f(R)$  gravity \cite{Bahamonde:2015hza, Bahamonde:2016cul, Bahamonde:2016grb, Capriolo2}.

One of the main advantages of studying TEGR and its extensions  is that these models can be easily recast as gauge theories allowing deep insights in the fundamental structure of gravity. In this perspective, quantizing TEGR and extensions could be extremely interesting  towards the realization of Quantum Gravity. Therefore, Quantum Cosmology could be useful to achieve  results possibly comparable with  observations.

Specifically, we worked out minisuperspace models by the Noether Symmetry Approach. 
We have also showed, by adopting suitable Lagrange multipliers, that Noether symmetries provide constraints on the form of the action that allow to simplify the dynamics and to find classical solutions. In particular, the Noether approach allows  to select  changes of variables such that the Wave Function of the Universe turns out to be peaked in  minisuperspaces containing  cyclic variables. In this case,  classical trajectories, representing observable universes  can be recovered. However, it is worth remarking that neither Noether system solutions nor the change of variables are unique. This means that  a careful choice is often important to recover exact useful solutions capable of describing  classical universes.

As the Wave Function of the Universe is related to the probability to get observable universes, the existence of Noether symmetries suggests when the Hartle criterion is working. Due to the experimental lack of many copies of the same universe, the resulting Wave Function is only \emph{related} to the probability to get a certain configuration, but it does not give  the whole probability amplitude itself. This shortcoming is also due to the lack of a self-consistent theory of Quantum Gravity.

In particular, we focused on $f(T)$, $F(T,\Box T)$,  $f(T,\G)$  and on  non-minimal coupled teleparallel models. The application of Noether Symmetry Approach to  the related point-like Lagrangians allows to select the functional form of the models. In all cases, the existence of symmetries permits to integrate the corresponding dynamical systems and to apply the Hartle criterion to find out observable universes. 

In a forthcoming paper, the method will be systematically confronted with observations to link the cosmological parameters with the Noether symmetries.

\section*{Acknowledgments}

The Authors acknowledge the support of {\it Istituto Nazionale di Fisica Nucleare} (INFN) ({\it iniziative specifiche} GINGER, MOONLIGHT2, QGSKY, and TEONGRAV). This paper is based upon work from COST action CA15117 (CANTATA), COST Action CA16104 (GWverse), and COST action CA18108 (QG-MM), supported by COST (European Cooperation in Science and Technology).

\appendix

\section{The Arnowitt-Dese- Misner formalism and Wheeler-DeWitt equation} 
\label{ADM}
In order to  understand the early time behavior of the Universe, we should treat it by using Quantum Field Theory, and represent its evolution by means of a Wave Function depending on the space-time variables. This is not fully possible due to the lack of a self-consistent theory of Quantum Gravity. 

However, adopting a canonical quantization approach and the related   ADM formalism allows to obtain the so called \emph{Wheeler-DeWitt} equation, which is a Schroedinger-like equation  whose solution, the Wave Function of the Universe, is a useful indication of probability amplitude for quantized superspace variables. 

In order to build up the formalism, let us start by considering the most general form of the Einstein-Hilbert action, namely \cite{Thiemann:2007zz, DeWitt:1967yk}
\begin{equation}
S = \frac{1}{2} \int_V \sqrt{-g}[R - 2 \lambda] d^4x + \int_{\partial V} \sqrt{h} K \; dx^3 \;,
\end{equation}
where $K = h^{ij} K_{ij}$ is the trace of the extrinsic curvature tensor of the three-dimensional hyper-surface $\partial V$ surrounded in the four-dimensional manifold, and $h$ is the determinant of the three-dimensional metric\footnote{In this appendix, $h_{ij}$ is the spatial component of the spacetime metric $g_{\mu\nu}$}. To get the Hamiltonian formulation, we perform a $(3+1)$-decomposition of the metric $g_{\mu \nu}$, so that the spatial components turn out to be the dynamical degrees of freedom. Choosing a set of coordinates $X^{\alpha}$, by means of coordinates foliation   transformation $X^\alpha \to X'^\alpha$, we can get  a family of  hypersurfaces  $x^0 = const$ where $x^i$ are the local coordinates on  each surface. 

In the four dimensions, to each time-like point $x^0$, it corresponds a space-like hypersurface $x^0 = k$, and the variation of $x^0$ provides the  required foliation. Moreover, any point of the hypersurface is labeled by a three-dimensional vectors basis $X^\alpha_i$, tangent to the surface itself. Since they are orthonormal to the surface vector $n^\nu$, the following two conditions hold:
\begin{equation}
g_{\mu \nu} X_i^\mu n^\nu = 0 \;\;\;\;\;\; g_{\mu \nu} n^\mu n^\nu = -1 \;.
\label{vettori normali e paralleli}
\end{equation}
It is also useful to define the deformation tensor as 
\begin{equation}
N^\alpha = \dot{X}^\alpha = \partial_0 X^\alpha(x^0, x^i)
\end{equation}
and, by decomposing $N\alpha$ in the previously defined basis of normal and tangent vectors as
\begin{equation}
N^\alpha = N n^\alpha + N^i X_i^\alpha \;,
\end{equation}
the metric tensor takes the form:
\begin{equation}
g_{\mu \nu} = \left(
\begin{matrix}

- (N^2 - N_i N^i) & N_j 
\\
N_j & h_{ij}.
\end{matrix}
\right)
\end{equation}
The scalar  $N$ and the vector $N^i$ are the so-called \emph{Lapse} and \emph{Shift} functions, respectively. With respect to these decompositions, the Lagrangian density turns out to be
\begin{equation}
\mathscr{L}=\frac{1}{2} \sqrt{h} N\left(K^{i j} K_{i j}-K^{2}+{ }^{(3)} R\right) +T.D.
\end{equation}
By defining the conjugate momenta as
\begin{eqnarray}
&& \pi \equiv \frac{\delta \mathscr{L}}{\delta \dot{N}}=0 \quad \pi^{i} \equiv \frac{\delta \mathscr{L}}{\delta \dot{N}_{i}}=0 \nonumber \\ 
&& \pi^{i j} \equiv \frac{\delta \mathscr{L}}{\delta \dot{h}_{i j}}=\frac{\sqrt{h}}{2}\left(K h^{i j}-K^{i j}\right)
\label{conjmom}
\end{eqnarray}
and the Hamiltonian density as 
\begin{equation}
\mathscr{H}=\pi^{i j} \dot{h}_{i j}-\mathscr{L},
\end{equation}
the commutation relations between the Hamiltonian $\mathcal{H}$ and the conjugate momenta $\pi$ are \cite{Capozziello:2012hm}:
\begin{equation}
\begin{cases}
\displaystyle \dot{\pi} = -\{\mathcal{H}, \pi \} = \frac{\delta \mathcal{H}}{\delta N} = 0
\\
\displaystyle \dot{\pi}^i = -\{\mathcal{H}, \pi^i \} =  \frac{\delta \mathcal{H}}{\delta N_i} = 0 \;.
\end{cases} 
\label{vincoli secondari sui momenti}
\end{equation}
With these definitions in mind, the first step for the  canonincal quantization  is to transform classical dynamical variables  to operators by  the appropriate commutation relations:
\begin{equation}
\begin{cases}
[\hat{h}_{ij}(x) , \hat{\pi}^{kl} (x')] = i \; \delta^{kl}_{ij} \; \delta^3 (x-x')
\\
\delta^{kl}_{ij} = \frac{1}{2} (\delta_i^k \delta_j^l + \delta_i^l \delta_j^k)
\\
[\hat{h}_{ij}, \hat{h}_{kl}] = 0
\\
[\hat{\pi}^{ij}, \hat{\pi}^{kl} ] = 0 \;,
\end{cases}
\end{equation}
Notice that the conjugate momenta are now promoted to operators, so that the definitions in Eq. \eqref{conjmom} take the form
\begin{equation}
\hat{\pi}=-i \frac{\delta}{\delta N} \quad \hat{\pi}^{i}=-i \frac{\delta}{\delta N_{i}} \quad \hat{\pi}^{i j}=-i \frac{\delta}{\delta h_{i j}}
\end{equation}
Furthermore, from Eq. \eqref{vincoli secondari sui momenti}, it automatically follows that the canonical quantization  imposes the constraint
\begin{equation}
\hat{{\cal{H}}}| \psi> = 0
\end{equation}
which leads to the Wheeler-De Witt equation, namely a Schr\"odinger-like equation of the form: 
\begin{equation}
\left(2 \; \nabla^2 - \frac{1}{2} \sqrt{h} \; ^{(3)} R \right)|\psi> = 0 \;.
\label{vincoli secondari}
\end{equation}
The function $\psi$ is defined on the configuration space of the 3-metrics, with $\psi \equiv \psi[h_{ij}(x)]$ describing the evolution of the gravitational field. The operator $\nabla^2$, is defined through the 3-dimensional metric as
\begin{equation}
\nabla^2 = \frac{1}{\sqrt{h}}\left(h_{i k} h_{j l}+h_{i l} h_{j k}-h_{i j} h_{k l}\right) \frac{\delta}{\delta h_{ij}}\frac{\delta}{\delta h_{kl}}.
\end{equation}

The space on which the Wave Function is defined is the space of all the possible 3-metrics, called \emph{superspace}. Since the Superspace is an infinite dimensional space, in order to solve the Wheeler-De Witt equation,  restrictions to  finite-dimensional \emph{minisuperspaces} are often needed. In such a way, the Wheeler-De Witt equation becomes a partial differential equation. It is worth noticing that  minisuperspaces represents \emph{toy models} which preserves some basic aspects of the entire theory, as some symmetry, but neglect others. From this point of view, the strongest hypothesis of Quantum Cosmology is to assume that these models are workable approximations of the complete theory. 

However, some difficulties in the interpretation of the Wave Function of the Universe  arises: in non-relativistic Quantum Mechanics, the dynamics of the system is described by the Schr\"odinger equation, so that the quantity $<\psi|\psi> \equiv \int \psi^* \psi \; dx^3$ is always positive and independent of time. The consequence of this fact is that   solutions of the Schr\"odinger equation constitute an infinite dimensional space, and the product $\psi^* \psi$ can be interpreted as the probability density to localize the particle into a give  configuration space. On the contrary, the Wheeler-DeWitt equation is similar to the Klein Gordon equation: it is possible to find a continuity equation, but the scalar product is not always positive-defined and then the probabilistic interpretation is missing.  

\section{The Noether Symmetry Approach\\}
\label{Noeth}
The Noether Symmetry Approach  allows to find out symmetry transformations and conserved quantities by means it is possible to reduce and solve dynamical systems.  In this appendix,  we outline the main properties of the Noether Approach  and point out  how to use  it in Quantum Cosmology.

Let ${\cal L}(t, \phi^i, \dot{\phi}^i)$ be a non-degenerate Lagrangian describing a dynamical system. An   infinitesimal transformation involving the fields $\phi^i$ and the coordinates is
\begin{equation}
\begin{cases}
\Lagr(t,\phi^i \dot{\phi}^i) \to \Lagr (\overline{t}, \overline{\phi}^i, \dot{\overline{\phi}}^i)
\\
\overline{t} = t + \epsilon \xi(t,\phi^i) + O(\epsilon^2)
\\
\overline{\phi}^i = \phi^i + \epsilon \eta^i(t,\phi^i) + O(\epsilon^2) \;.
\end{cases}
\label{trasformazioni}
\end{equation}
The generator of the transformation can be written as:
\begin{equation}
X^{[1]} =  \xi \frac{\partial }{\partial t} + \eta^i \frac{\partial }{\partial \phi^i} + (\dot{\eta}^i - \dot{\phi}^i \dot{\xi}) \frac{\partial}{\partial \dot{\phi}^i}.
\label{prolungamento vettore noether}
\end{equation}
Noether's first theorem states that if the relation 
\begin{equation}
X^{[1]} \Lagr + \dot{\xi} \Lagr = \dot{g}
\label{theorem}
\end{equation}
holds, then the quantity 
\begin{equation}
I(t,\phi^i,\dot{\phi}^i) = 	\displaystyle \xi \left(\dot{\phi}^i \frac{\partial \Lagr}{\partial \dot{\phi}^i} - \Lagr \right) - \eta^i \frac{\partial \Lagr}{\partial \dot{\phi}^i} + g(t,\phi^i) \, ,
\label{QC}
\end{equation}
is a constant of motion. The function $\xi$ depends on all the variables of the configuration space  and represents the infinitesimal generator related to the change of coordinates; $\eta^i$ is, in turn, related to the fields variation. The function $g(t,\phi^i)$ is a four divergence whose value does not affect the dynamics of the system. This is the general version of Noether's theorem holding both for internal and  external symmetries. A   particular case can be obtained  by the condition $\xi = 0$. In the applications to teleparallel Lagrangians discussed in this work, we only considered internal symmetries, where the  Noether vector is written as:
\begin{equation}
X = \eta^i \frac{\partial }{\partial \phi^i} + \dot{\eta}^i  \frac{\partial}{\partial \dot{\phi}^i} 
\label{generatore}
\end{equation}
and the identity \eqref{theorem}  reduces to $X \Lagr = 0$ after imposing $g=0$. This means that, if the transformation does not involve coordinates, internal symmetries arise if the Lie derivative of the Lagrangian along the flux of X vanishes:
\begin{equation}
L_X {\cal L} = X {\cal L} = \eta^i \frac{\partial {\cal L}}{\partial \phi^i} + \dot{\eta}^i \frac{\partial {\cal L}}{\partial \dot{\phi}^i}
\end{equation}
and phase flux is conserved along X. Therefore, setting to zero the Lie derivative of Lagrangian it is possible to find the symmetries of the theory and the corresponding conserved quantity, that is
\begin{equation}
\Sigma_0 = \eta^i \frac{\partial \Lagr}{\partial \dot{\phi}^i}\,.
\label{sigma0}
\end{equation} 
Notice that Eq. \eqref{QC} reduces to Eq. \eqref{sigma0} when $\xi = 0$. 

In Quantum Cosmology, the configuration space is the minisuperspace and,
 to find the Hamiltonian and to solve the Wheeler-DeWitt equation, cyclic variables are identified  by an appropriate change of coordinates. A possible transformation is such that the conjugate momentum of the new variable $z$ coincides with the conserved quantity $\Sigma_0$ provided by the Noether symmetry, that is:  
\begin{equation}
\pi_z = \Sigma_0 \;.
\end{equation}
According to Eq. \eqref{sigma0}, the above relation can be written as:
\begin{equation}
\left(\eta^i\frac{\partial {\cal L}}{\partial \dot{\phi}^i} \right) = \frac{\partial {\cal L}}{\partial \dot{z}} \;,
\label{momento zeta}
\end{equation}
In this way, also the generator of the symmetry must transform accordingly; let $X'$ be the generator written in terms of the new variables:
\begin{equation}
X' = \eta'^i \frac{\partial}{\partial \overline{\phi}^i} + \dot{\eta}'^i \frac{\partial}{\partial \overline{\dot{\phi}}^i}
\end{equation}
Setting the component $\eta'^1$ of the new infinitesimal generator $\eta'^i$ equal to 1 and $\eta'^j$ (with $j \neq 1$) equal to zero, the conserved quantity \eqref{sigma0},  corresponding  to the variable $\overline{\phi}^1$, turns out to be equal to the conjugate momentum $\pi_{\phi^1}$. More formally, let us consider the coordinates transformation $\phi^i \to \Phi(\phi^i)$ and the corresponding \emph{inner derivative}
\begin{equation} 
i_X d\Phi \equiv \dot{\eta}'^i \frac{\partial \Phi}{\partial \phi^i} \;.
\end{equation}
The new generator can be written in terms of the inner derivative as:
\begin{equation}
X' =  (i_X d\Phi^k) \frac{\partial}{\partial \Phi^k} + \frac{d}{dt}(i_X d\Phi^k)\frac{\partial}{\partial \dot{\Phi}^k}
\label{generatore inner}
\end{equation}
so that, by setting the first component of the new infinitesimal generator equal to 1 and the others equal to zero, we obtain
\begin{equation}
i_X d\Phi^1 = \eta'^i \frac{\partial \Phi^1}{\partial \phi^i} = 1 \;\;\;\;\;\;\; i_X d\Phi^j = \eta'^i \frac{\partial \Phi^j}{\partial \Phi^i} = 0 \;\;\;\;\;\;\; j \neq 1\;.
\label{relzioni per cambio variabili}
\end{equation}
The conditions \eqref{relzioni per cambio variabili} allows to introduce a cyclic variable by means of a methodical procedure. This procedure, after quantizing the theory and getting the Wave Function of the Universe, points out when the  Hartle criterion holds: conserved momenta correspond to oscillating components of the wave function. In this case the Hamilton-Jacobi equations can be exactly integrated giving classical trajectories. According to the Hartle criterion, such trajectories correspond to observable universes. 
The same procedure does not hold for the extended theorem, whose conserved quantities cannot be replaced by the conjugate momenta of given variables. For this reason, external symmetries have not been  taken into account.

From the Quantum Cosmology   point of view, all the variables (and symmetries) are internal to the system and then the Noether theorem can be restricted only to internal variables.


\end{document}